\documentclass[12pt]{iopart}

\usepackage{iopams,graphicx}  

\def\p{\partial}
\def\bn{\textbf{n}}
\def\be{\begin{equation}}
\def\ee{\end{equation}}

\begin{document}

\title{Saddle-splay elasticity and field induced soliton in~nematics}




\author{O V Manyuhina}

\address{Nordita, Royal Institute of Technology \& Stockholm University,  Roslagstullsbacken~23, SE-10691 Stockholm, Sweden}
\ead{oksanam@nordita.org}

\begin{abstract}

The symmetry breaking Freedericksz transitions, when a uniformly aligned nematic state is replaced by a homogeneously or periodically distorted state, have been extensively studied before. Here we analyse the influence of the saddle-splay elasticity on the non-linear ground state of nematic liquid crystal in presence of magnetic field above the Freedericksz threshold. We identify the bifurcation point when the localised soliton-like state is linearly unstable with respect to the perturbations of the wavevector in the direction perpendicular to the initial plane of soliton. This instability occurs only if the ratio of the saddle-splay elastic constant to the elastic modulus of nematics in the one-constant approximation is above the critical value  $|K_{24}/K|\geqslant0.707$.

\end{abstract}

\maketitle

\section{Introduction} 

The nematic liquid crystal is anisotropic fluid, composed of rod-like molecules, characterised by a unit vector  $\bn$ called the director,  pointing along the averaged orientation of molecules~\cite{degennes:book}. Since the pioneering work of Zolina and Freedericksz~\cite{freed} the field induced Freedericksz transitions is one of the most studied and most useful phenomenon in physics of liquid crystals.  When the magnetic (or electric) field is applied  perpendicular to the uniformly aligned nematic,  confined between two parallel plates,  above a certain threshold the director experiences deformations and tends to align along the field. The critical threshold is inversely proportional to the thickness of the nematic sample $d$, and quantifies the competition between the magnetic field $B$ and the boundary effects mediated by elastic forces, given by
\be\label{eq:Bc}
B_c=\frac \pi d \sqrt{\frac K{\chi_a}}, 
\ee
where $\chi_a$ is the magnetic susceptibility  and  $K$ is the elastic constant. The symmetry breaking in classical Freedericksz transitions happens in one direction along the thickness of the nematic  sample and the observed  distortions can be explained within a continuum theory of liquid crystals~\cite{stewart:book}. The discovery of a periodic  splay-twist Freedericksz transitions by Lonberg and Meyer~\cite{lonberg}, when a homogeneous distortions are replaced by a spatially periodic pattern of stripes at even lower threshold value than $B_c$,  posed new questions on the equilibrium structure of the nematic in presence of the magnetic field. Taking into account more realistic boundary conditions,  namely a finite anchoring strength~\cite{gaetano:anch} and the saddle-splay elastic term~\cite{gaetano:K24},  would influence the critical threshold and result into the periodic saddle-splay Freedericksz transitions~\cite{kralj}. 


The problem of finding the critical threshold in different classes of Freedericksz transitions was formulated in terms of linear  perturbation theory, assuming a uniformly aligned  ground state. In this paper we consider the nematic liquid crystal in presence of high magnetic fields $B\gg B_c$, when the ground state is  initially distorted,  more precisely, in the bulk  the director is aligned parallel to the field, and in the boundary layer at the surfaces the director  reorients rapidly. The equilibrium configuration found within the boundary layer is the kink-like soliton,  because magnetic energy enters the Lagrangian of the system in a non-linear way. To our knowledge, the influence of the saddle-splay elastic constant on this non-trivial localised  ground state in presence of high magnetic field $B\gg B_c$ was not considered theoretically before. In essence the boundary layer can be conceived as  thin nematic film   spread on liquid substrate, subjected to the antagonistic boundary conditions~\cite{cazabat:2011}.  Nevertheless, in this case the ground state  depends linearly on the coordinate along the thickness, so called hybrid aligned nematic, because there is no non-linearity entering the free energy. It has been shown that the saddle-splay term plays an important role for thin nematic films with weak anchoring conditions, when the instability towards a periodically distorted  stripe phase was found~\cite{sparav:1995,LP:1995,our:epl}.

In the present paper we treat the problem within the Oseen--Zocher--Frank continuum theory of liquid crystals, assuming the one-constant approximation of the elastic free energy~\cite{degennes:book}. Starting from the non-linear ground state of the nematic director in presence of high magnetic field, we consider the onset of instability towards periodically deformed state, with stripes being parallel to the direction of the field. As a result, the stripe phase can be energetically favoured if  the absolute value of the saddle-splay elastic constant is higher than  $K/\sqrt{2}$. The found analytic expression for the critical wavenumber as function of the control parameter suggests  new experimental way to identify simultaneously the anchoring strength and  the saddle-splay elastic constant.

\section{\label{sec:kink}Soliton solution}

\begin{figure}[h]
\centering
\includegraphics[width=0.45\linewidth]{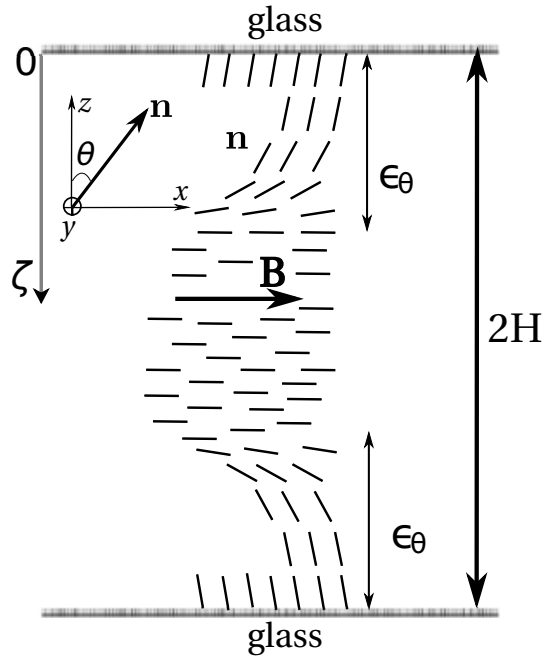}
\caption{\label{fig:cell}Schematic representation of the director $\bn\leftrightarrow-\bn$ configuration in presence of magnetic field $B$ in a cell of $2H$ thickness.  In the boundary layer with dimensionless thickness $\epsilon_\theta \ll 1$~(\ref{eq:eps1}), $\bn$ changes from the orientation parallel to the field, along $x$-axis, towards the one imposed by the glass surface through the boundary condition~(\ref{eq:bc}). The equilibrium angle $\theta$ (\ref{eq:theta}), between the director $\bn$ and the $z$-axis is function of the scaled coordinate $\zeta$,  where $\zeta\to\infty$ corresponds to the inner border of the boundary layer.}
\end{figure}

Let us consider nematic liquid crystal constrained between two glass plates, separated by distance $2H$ along $z$-axis, and subjected to the magnetic field $B$  parallel to the $x$-axis as shown in Fig.~\ref{fig:cell}. We assume that glass surfaces impose a homeotropic anchoring, favouring director alignment  along the normal to the surface,   and the positive magnetic susceptibility of molecules $\chi_a$, so that $\bn$ tends to align along the field. Above the Freedericksz threshold  $B\gg B_c$ (\ref{eq:Bc}) we get $\bn$ parallel to the field  in the bulk, and two regions in the vicinity of glass surfaces, known as boundary layers, where the reorientation of $\bn$ takes place.  The equilibrium configurations of nematics minimize the total free energy, written as the sum of the Frank free energy  in the one-constant approximation and the magnetic energy
\be\label{eq:fb}
{\cal F}_b=\frac 12 \int_V dV\,\big\{K |\nabla \bn|^2 - K_{24}\nabla\cdot [ (\nabla \cdot \bn)\bn+\bn\times\nabla\times\bn] - B^2\chi_a (\bn\cdot{\bf e}_x)^2\big\}.
\ee
The second divergent term, where  $K_{24}$\footnote{Note that this is a `pure' saddle-splay elastic constant, where the contributions from splay, twist and bend terms in the one-constant approximation have already been taken into account by using the following identity
$$
(\nabla\cdot \bn)^2+(\bn\cdot\nabla\times \bn)^2 + |\bn\times\nabla\times\bn|^2\equiv|\nabla\bn|^2+\nabla\cdot[(\nabla\cdot \bn)\bn +\bn\times\nabla\times\bn].
$$ } is the saddle-splay elastic constant, can be transformed into a surface integral,  entering boundary conditions in the case of non-planar geometry. Moreover, at the glass surface we introduce the additional energy associated with the anchoring, which we write in the Rapini--Papoular form as
\be
{\cal F}_s=\frac 12 \int_S dS\,\big\{W_0-W_a(\bn\cdot {\bf e}_z)^2\big\}, 
\ee
where $W_a>0$ is the anchoring strength and the sign minus reflects that  anchoring favours director alignment parallel to $e_z$.

We are looking first for the equilibrium configuration of the nematic director in 2D case with the following parametrization $\bn = \sin\theta(z){\bf e}_x+\cos\theta(z){\bf e}_z$, so that $|\bn|^2=1$ and $\theta$ is an angle which director makes with $z$-axis. Introducing the dimensionless coherence length and the scaled coordinate $\zeta$
\be\label{eq:eps1}
\epsilon_\theta = \frac 1 H\sqrt{\frac K{\chi_aB^2}}, \qquad \frac z H  =1 -\epsilon_\theta \zeta,
\ee
the Euler--Lagrange equation for $\theta$ associated with  (\ref{eq:fb}) takes the form
\be \label{eq:theta0}
\p_{\zeta\zeta} \theta + \sin \theta\cos\theta=0.
\ee
Without loss of generality we consider the upper half-plane and the boundary conditions  in the bulk ($\zeta\to\infty$) and at the glass surface ($\zeta\to0$) are respectively
\be\label{eq:bc}
 \theta|_{\zeta\to\infty}= \frac \pi2,\quad \p_\zeta\theta|_{\zeta\to \infty}= 0,\qquad \rho(\p_\zeta\theta)|_{\zeta=0}=(\sin\theta\cos\theta)|_{\zeta=0},
\ee
where  $\rho \equiv \sqrt{K B^2\chi_a}/W_a$ is the dimensionless parameter relating the strength of the magnetic field to the anchoring.
The first integral of (\ref{eq:theta0}), satisfying (\ref{eq:bc}) in the bulk, is 
\be
\p_\zeta\theta = \cos \theta,
\ee
which can be integrated again yielding
\be\label{eq:theta}
\theta (\zeta)= \arcsin \bigg(\frac {Ae^{2\zeta} -1}{Ae^{2\zeta}+1}\bigg), 
\ee
where the integration constant $A$ satisfies the boundary condition at the interface (\ref{eq:bc}), yielding
\be\label{eq:Arho}
A(\rho) = \frac{1+\rho}{1- \rho}.
\ee
In order to have a non-trivial solution $\rho\leqslant 1$ should hold. In the case of infinite anchoring $W_a\to\infty$ ($\rho\to0$) we get $A=1$ and thus $\theta(\zeta)=\arcsin(\tanh \zeta)$. The obtained soliton-like solution is due to non-linearity in (\ref{eq:theta0}), resulting in the localisation of the distortion of~$\bn$ within the boundary layer of thickness $\epsilon_\theta H$.

\section{\label{sec:linear}Linear stability analysis}

Let us consider a small perturbation of the director in $yz$-plane
\be\label{eq:bn}
\bn= \sin(\theta+\psi)\cos\phi\,{\bf e}_x+\sin(\theta+\psi)\sin\phi\,{\bf e}_y+\cos(\theta+\psi){\bf e}_z,
\ee
 where $\phi(y,z)$ and $\psi(y,z)$ are assumed to be small $O(\varepsilon)$ and periodic functions with respect to the variable $y$.  Then, expanding  the free energy density up to $O(\varepsilon^2)$  we find the perturbed bulk $f_b^{(2)}$ and the surface $f_s^{(2)}$ contributions, respectively 
\begin{eqnarray}
f_b^{(2)}& =  \frac K2 \big[(\p_z\psi)^2 +(\p_y\psi)^2+\sin^2\theta((\p_z\phi)^2+(\p_y\phi)^2)\big]-\nonumber \\ &\qquad -\frac{B^2\chi_a}2(\cos2\theta \psi^2-\sin^2\theta\phi^2),\label{eq:fb2}\\
f_s^{(2)}& = \frac {W_a}2(\cos2\theta\,\psi^2)\big|_{\zeta=0} -K_{24} (\sin^2\theta \,\phi\,\p_y\psi)\big|_{\zeta=0},\label{eq:fs2}
\end{eqnarray}
where the saddle-splay term favours non-zero distortions of $\phi$ and $\psi$. To characterise the instability of the planar solution (\ref{eq:theta}) with respect to the perturbation of wavevector $q$ in $y$-direction we are searching for the periodic solution in the form 
\be\label{eq:fg}
\phi(z,y)=g(z)\cos(qy), \qquad \psi(z,y)=f(z)\sin(q y).
\ee
The variational problem for $f$ and $g$ associated with (\ref{eq:fb2}), given the solution of $\theta(\zeta)$ (\ref{eq:theta}), leads to the following differential equations  
\begin{eqnarray}
\p_{\zeta\zeta} f-f\bigg(\omega^2-\frac{8A e^{2\zeta}}{(Ae^{2\zeta}+1)^2}\bigg)=&0,\\
\p_{\zeta\zeta}g+\p_{\zeta}g\frac{8A e^{2\zeta}}{A^2e^{4\zeta}-1}-g \omega^2=&0,
\end{eqnarray}
written in terms of the scaled variable $\zeta$ (\ref{eq:eps1}) and another dimensionless variable $\omega^2\equiv 1+q^2\epsilon_\theta^2H^2$ is introduced.
The resulting solutions, vanishing in the bulk ($\zeta\to\infty$) can be cast into the  form
\begin{eqnarray}
f(\zeta)&=C_1 \frac{e^{-\zeta\omega}(Ae^{2\zeta}(1+\omega)+\omega-1)}{1+Ae^{2\zeta}}, \label{eq:f}\\
g(\zeta)&=C_2\frac{A^{(1-\omega)}e^{-\zeta\omega}(Ae^{2\zeta}(1+\omega)+\omega-1)}{(\omega-1)(Ae^{2\zeta}-1)}\label{eq:g},
\end{eqnarray}
where the integration constants $C_1$ and $C_2$ should satisfy the following boundary conditions at the glass surface $\zeta=0$
\begin{eqnarray}
\p_\zeta f (0) &=\bigg(\frac1 \rho -2 \rho \bigg) f(0) -\tau \sqrt{\omega^2-1}\, g(0),\\ 
\p_\zeta g(0) &=-\tau \sqrt{\omega^2-1} f (0), \qquad \tau\equiv\frac{K_{24}}K.
\end{eqnarray}
The resulting system of linear equations $\sum {\cal M}_{ij}C_j=0$ with respect to unknowns $C_1$ and $C_2$ has a non-trivial solution  if and only if  the determinant of the $2\times2$ matrix of the coefficients is zero, $\det {\cal M}=0$.   This condition yields the implicit relationship between the dimensionless parameters of the system $\rho,\omega$ and $\tau$, the latter should satisfy Ericksen's inequalities for nematic liquid crystals~\cite{ericksen:1966}, which in the one-constant approximation reduce to   $|\tau|\leqslant 1$. If $\det {\cal M}>0$ the 2D non-linear state (\ref{eq:theta}) is stable ($C_i=0$), otherwise an instability towards the periodically distorted state occurs, with a wavevector $q$ determined at the bifurcation point $\det {\cal M}=0$. In the following section we analyse the  instability threshold and identify the associated critical parameters. 


\section{\label{sec:periodic}Periodic solution}


\begin{figure}[h]
\centering
\raisebox{6.5cm}{a)}\includegraphics[width=0.45\linewidth]{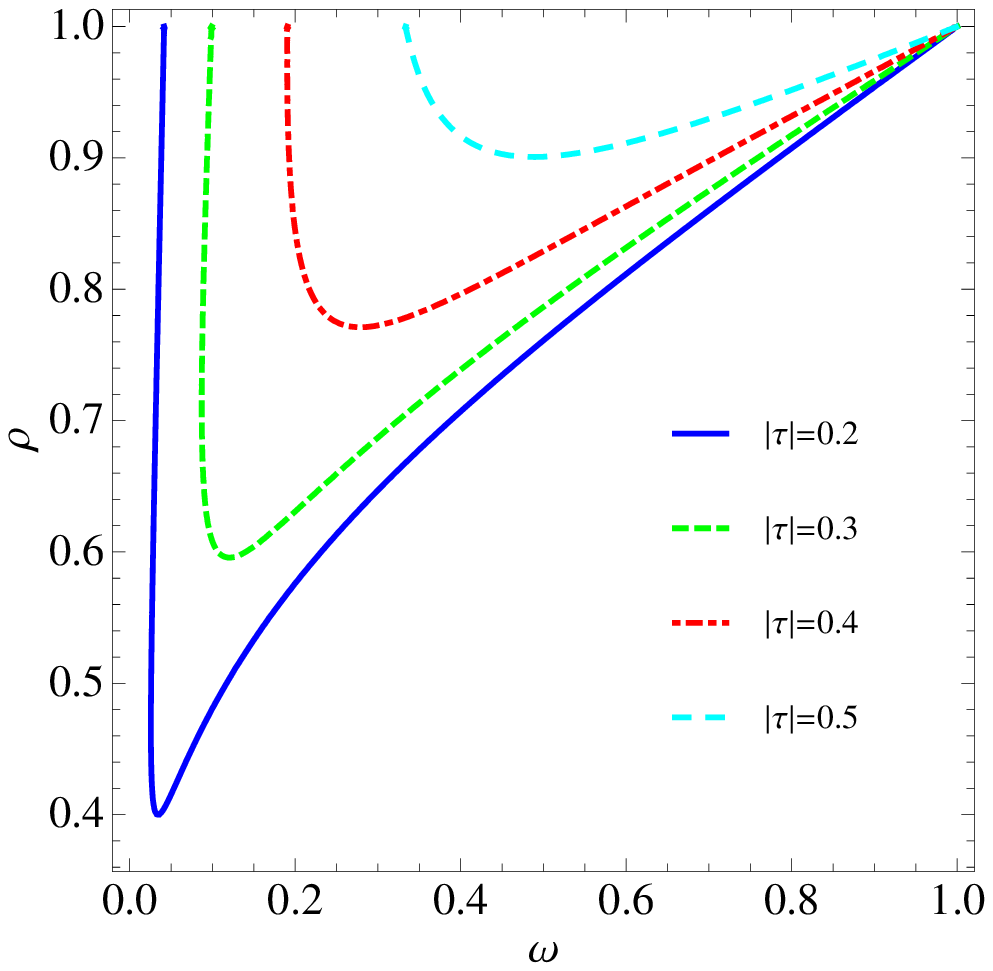}\hfill
\raisebox{6.5cm}{b)}\includegraphics[width=0.45\linewidth]{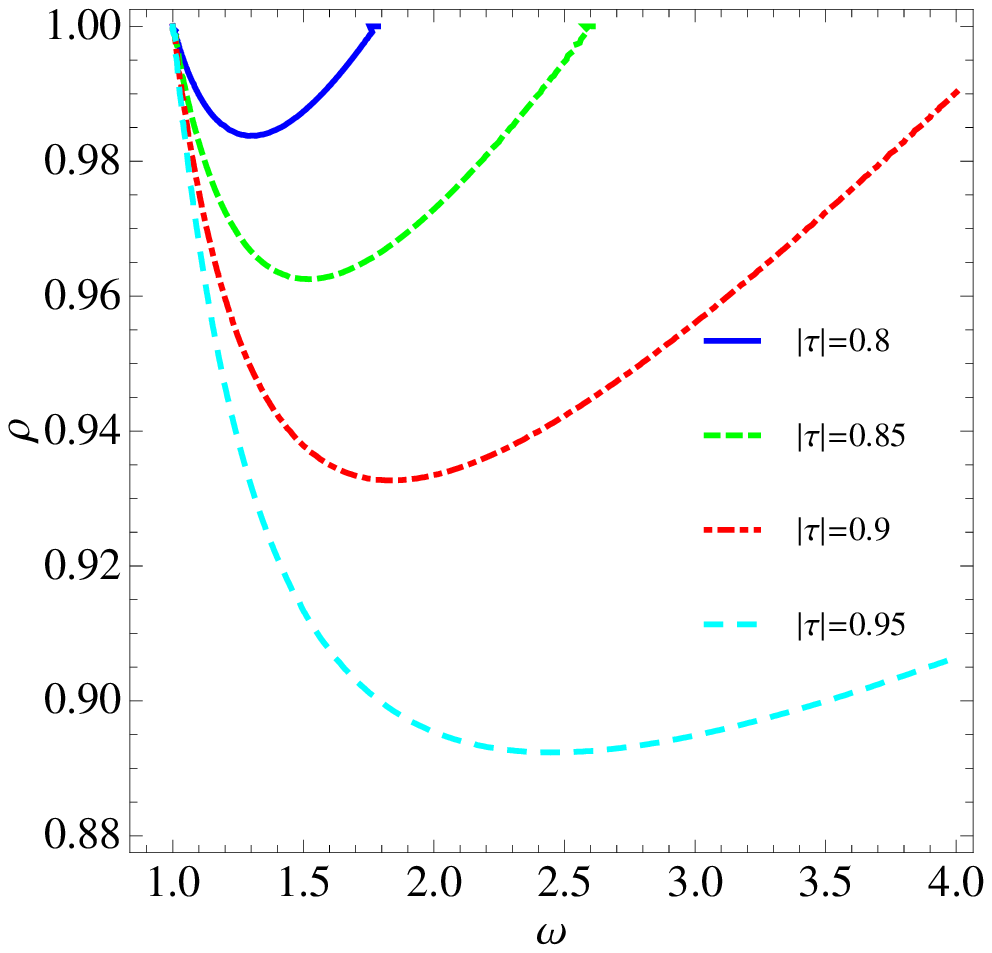}
\caption{\label{fig:det}The solution of the equation $\det{\cal M}=0$ (\ref{eq:detM}) for different values of the saddle-splay constant, given by $\tau=K_{24}/K$: (a) $|\tau|<\tau_c$, (b)  $|\tau|>\tau_c$ with $\tau_c=1/\sqrt{2}$. The minima of the curves in (b) define the critical wavenumbers $q_c$ related to $\omega$, which occur at the critical ratio of the magnetic energy to the anchoring  energy, given by~$\rho_c$. }
\end{figure}

We make use of {\it Mathematica~7} to compute the determinant of the matrix 
\begin{eqnarray}
\det{\cal M} &=\frac {4 (1 +\rho)^2 (\frac{1 +\rho}{1 -\rho})^{-2 \omega} (\rho + \omega)^2 }{(1-\rho)^2\rho^4 (\omega-1)^2} \Big[\omega^2(1 +  2\rho \omega)-\rho^2 (\tau^2-1) \omega^2 (\omega^2-1) \nonumber\\&\quad -\rho^3\omega (1 - 2\tau^2  + (1 +  2\tau^2) \omega^2) -\rho^4 (\omega^2(1 +\tau^2)-\tau^2)\Big],\label{eq:detM}
\end{eqnarray}
and to plot  the  curves $\det {\cal M}=0$  in the $\omega$-$\rho$ parameter space for a given value of $\tau$,  as shown in  Fig.~\ref{fig:det}. The minimum of every curve gives the critical values of the governing parameter $\rho_c$, characterising the relative strength of magnetic field compared to the anchoring of the surface, and the corresponding scaled wavenumber $\omega_c$ of the perturbed state. We notice that not all the considered values of the saddle-splay elastic constant, satisfying Ericksen's inequalities, result in the real value for the  critical wavenumber. In Fig.~\ref{fig:det}a  $\omega_c<1$ and the corresponding wavenumber $q_c= \sqrt{\omega_c^2-1}/(\epsilon_\theta H)$ is complex and physically irrelevant, therefore, the periodically modulated state is not feasible. On the contrary, in Fig.~\ref{fig:det}b the curves reach their minimum at $\omega_c>1$, corresponding to real values of $q_c$, thus dividing the parameter space into two regions: i) below the curve ($\rho<\rho_c$) the 2D  base state (\ref{eq:theta}) is stable ($\det {\cal M}>0$), ii) above $\rho_c$ the periodically modulated state is energetically preferred one. The underlying picture can be understood in terms of the control parameter $|\tau|$, which being below its critical value $\tau_c$ results in the stability of the unperturbed 2D state and above $\tau_c$ one may find a supercritical bifurcation to the periodically distorted state with non-zero real~$q_c$. 

\begin{figure}[h]
\centering
\raisebox{4.5cm}{a)}\includegraphics[width=0.45\linewidth]{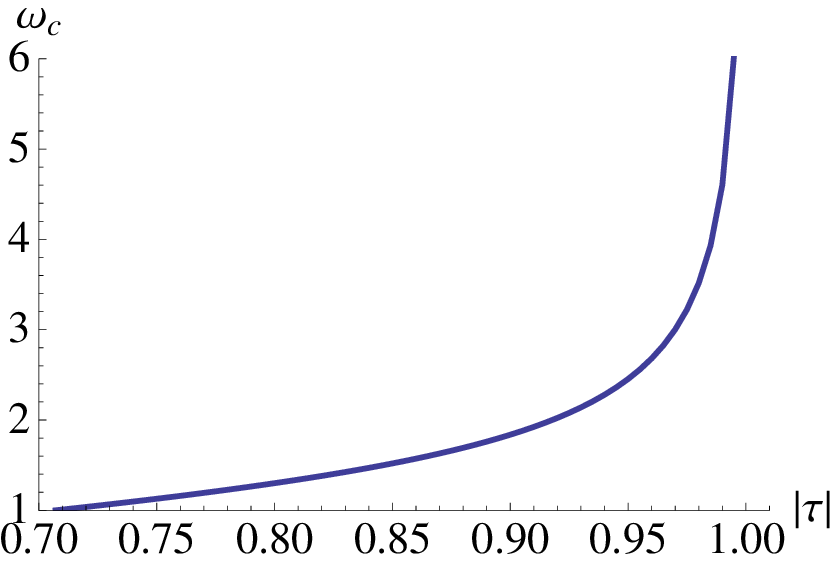}\hfill
\raisebox{4.5cm}{b)}\includegraphics[width=0.45\linewidth]{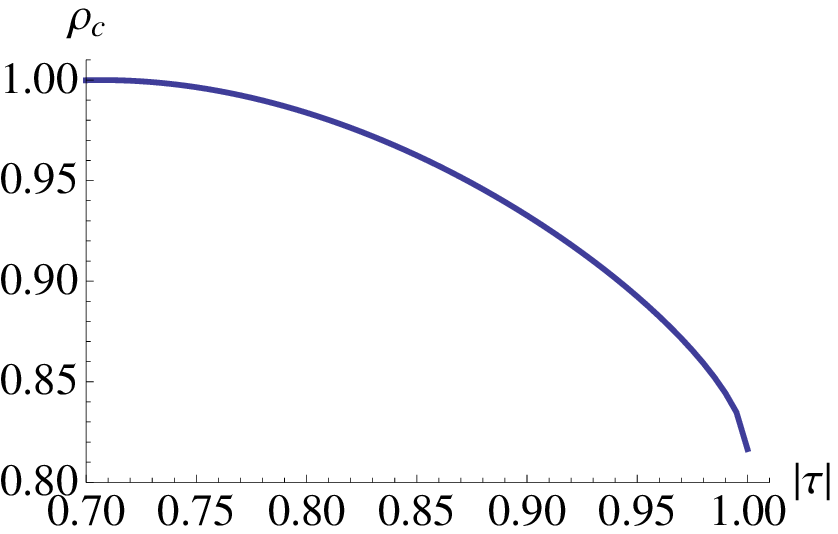}
\caption{\label{fig:crit}The critical parameters $\omega_c^2= 1+q_c^2\epsilon_\theta^2H^2$ and $\rho_c = \sqrt{K B^2\chi_a}/W_a$ at the instability threshold  as function of $\tau= K_{24}/K$, satisfying Ericksen's inequalities.}
\end{figure}

Let us plot in Fig.~\ref{fig:crit} the critical parameters $\omega_c$ and $\rho_c$ at the bifurcation point  as function of the saddle-splay elastic constant $\tau$.  The curves show that the perturbation with a finite wavelength, when $\omega_c>1$, appears only for $|\tau|> 0.7$. The critical value of $\tau$ is defined by taking the  following limit $\lim_{{\rho\to1, \omega\to 1}}$ of (\ref{eq:detM}), which exists if and only if $|\tau_c|=1/\sqrt{2}\simeq0.7$.  For the saddle-splay elastic constant, which does not satisfy  Ericksen's inequalities, namely $|\tau|> 1$ there is no critical point. Requiring the determinant (\ref{eq:detM}) to vanish in the limit $|\tau|\to1$ and $\omega\to\infty$, we find $\rho_c=\sqrt{2/3}\simeq 0.82$. Therefore, the instability threshold between the soliton-like ground state (\ref{eq:theta}) and the periodically distorted state (\ref{eq:fg}) with finite wavelength  exists for the following range of parameters  $1/\sqrt{2}<|\tau|<1$ and $1>\rho_c>\sqrt{2/3}$.

\begin{figure}[h]
\centering
\includegraphics[width=0.95\linewidth]{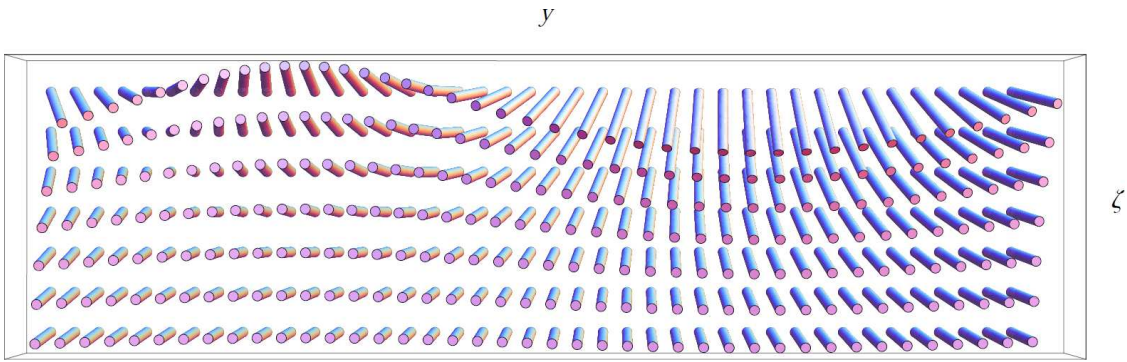}
\caption{\label{fig:director}The illustration of the space variation of the director ${\bf n}$~(\ref{eq:bn}) in $\zeta-y$ plane above the threshold for the formation of stripes. The boundary layer is similar to the one in Fig.~\ref{fig:cell}. In the bulk (bottom of the box) the director is aligned along the magnetic field parallel to the $x$-axis (not shown), while in the vicinity of the glass surface (top of the box) the periodic distortions of the director along $y$-axis reach maximum. The linear stability analysis does not allow to find the amplitude of the perturbation which is assumed to be $O(1)$. The chosen values of parameters are $\rho=0.95$, which gives $W_a\simeq 2\cdot 10^{-5}$~J/m$^2$ for $B=7$~T, the wavelength $\lambda \simeq2~\mu$m corresponds to $\omega=1.5$ and the saddle-splay elastic constant is assumed $|\tau|=0.9>0.88$. According to Fig.~\ref{fig:det}b we are slightly above the threshold.}
\end{figure}

According to the plotted critical curves, for the magnetic field of the order of Tesla  we may find the periodicity of stripes of the order of microns if the anchoring strength  $W_a\sim 10^{-5}$~J/m$^2$. These values are experimentally accessible.  Therefore,  if we know the strength of magnetic field  and can measure the periodicity of stripes, we can identify the saddle-splay elastic constant $\tau$ from   Fig.~\ref{fig:crit}a,  and consequently estimate the anchoring strength $W_a$ from  Fig.~\ref{fig:crit}b. The strong homeotropic anchoring  ($\rho<1$) leads to smaller wavelength of distorted stripe state, compared to the weak  anchoring ($\rho\simeq1$), yielding the long-wavelength periodic pattern. In Fig.~\ref{fig:director} we show the equilibrium nematic state with periodic distortions along $y$-axis for  typical values of the parameters $\rho, \omega, \tau$ slightly above the critical threshold. The resulting stripes are formed  in the vicinity of the glass surface  along the direction of the applied magnetic field or $x$-axis, which is not shown.

\section{Conclusions}

The intrinsic anisotropy of nematic liquid crystals leads to a non-linear elastic response  to the applied magnetic field, and therefore the equilibrium state is described by a localised soliton (\ref{eq:theta}). Nevertheless, this localised ground state can become unstable with respect to the perturbations of the wavevector $q_c$, when  the saddle-splay elastic term becomes important. Within the linear stability analysis we examined the onset of stripe instability, arising due to the interplay between the elastic and magnetic forces in the bulk versus the anchoring and saddle-splay forces, favouring the undulations at  the surface.  The bifurcation to the periodically deformed stripe state with a finite wavelength happens only if $1/\sqrt{2}<|\tau|<1$. Moreover, the threshold is characterised by the critical ratio of the magnetic energy to the anchoring energy, which falls into the range $\sqrt{2/3}<\rho_c<1$. If we know the parameters such as anchoring at the glass surface together with the saddle-splay and elastic moduli of nematic,  we can estimate, assuming the one-constant approximation,  the critical value of the magnetic field and the corresponding critical wavelength of stripes. On the other hand, varying the magnetic field and observing  the periodic deformations of nematic in the vicinity of the surface, we can identify simultaneously the anchoring strength and the poorly studied saddle-splay elastic constant, by  comparing experimental data with theoretical predictions (\ref{eq:detM}) based on  the  analytic solutions. To account for the difference in splay, twist and bend elastic constants would require  numerical analysis for solving non-linear partial differential equations. We believe that considering ground state different from the widely explored planar nematic is important for analysing experimental observations. Moreover, studying  competing interactions  between the surface effects and high magnetic fields above the Freedericksz threshold may give insight into fundamental  physics of liquid crystals as well as contribute to practical applications.  

\ack 
It is a pleasure to acknowledge valuable discussions with Gaetano Napoli.


\end{document}